# A Low-Cost Monopulse Receiver with Enhanced Estimation Accuracy Via Deep Neural Network

Hanxiang Zhang, *Student Member, IEEE*, Saeed Zolfaghary Pour, *Student Member IEEE,* Hao Yan, *Student Member, IEEE*, Powei Liu*, Student Member, IEEE*, Bayaner Arigong, *Senior Member, IEEE*

*Abstract*—In this paper, a low-cost monopulse receiver with an enhanced direction of arrival (DoA) estimation accuracy via deep neural network (DNN) is proposed. The entire system is composed of a 4-element patch array, a fully planar symmetrical monopulse comparator network, and a down conversion link. Unlike the conventional design topology, the proposed monopulse comparator network is configured by four novel port-transformation rat-race couplers. In specific, the proposed coupler is designed to symmetrically allocate the sum (Σ) / delta (Δ) ports with input ports, where a 360° phase delay crossover is designed to transform the unsymmetrical ports in the conventional rat-race coupler. This new rat-race coupler resolves the issues in conventional monopulse receiver comparator network design using multilayer and expensive fabrication technology. To verify the design theory, a prototype of the proposed planar monopulse comparator network operating at 2 GHz is designed, simulated, and measured. In addition, the monopulse radiation patterns and direction of arrival are also decently evaluated. To further boost the accuracy of angular information, a deep neural network is introduced to map the misaligned target angular positions in the measurement to the actual physical location under detection.

## I. INTRODUCTION

MONOPULSE tracking radar encodes the RF signals to extract the accurate angular information from a target under detection. It has been widely applied on tracking, short-range sensing, image processing, medical diagnostics, low cost IoT, and autonomous driving [1]-[3]. The signals received from the monopulse antenna array are preprocessed through an analog comparator feeding network directly at RF waveform domain to simultaneously obtain the coordinate position along two spatial dimensions and velocity of a moving object. The comparator network is often composed by four identical 180° couplers, and it can be categorized into the unsymmetrical and symmetrical circuit topology. For unsymmetric topology of comparator network, it is usually composed of conventional microstrip line rat-race coupler or 90° couplers with 90° phase shifters. For examples, in [4]-[5], four rat-race couplers are applied to construct comparator network, and multi-layer PCB with back feeding are required to feed the antenna array. To resolve the back feeding, in [6]-[8], a combination of four 90° couplers with 90° phase shifters are applied to design the comparator network, which often lead to a limited operating bandwidth and transmission imbalance.

Comparing with unsymmetrical topology, the symmetrical comparator network shows significant advantages as layout symmetric for input and output ports, low-cost fabrication, small amplitude and phase distortion, which are the key design parameters for obtaining high accuracy angular information in monopulse radar system. To realize symmetric topology, many design methodologies are investigated in the past including new components design and new fabrication process etc. For examples, in [9], a symmetrical port 180° coupler is realized by combining in-phase power splitter and Marchand balun, and the air bridge is applied to connect the two stage couplers. In [10]-[11], a multilayer microstrip structure is designed to integrate the antenna array with the conventional comparator network. In [12]-[15], the monopulse arrays are designed using substrate integrated waveguide (SIW), irregular interconnected waveguide, and magic-Tee, respectively. In [16], a 2D conical monopulse antenna has been proposed, and a planar symmetrical 180° coupler without ports crossing is developed in [17] to design symmetrical comparator network, where two transmission lines interconnect two hybrid couplers. Overall, the approaches mentioned above have tried to resolve the layout difficulty of 180° coupler in a planar comparator network, where the sum (Σ) and delta (Δ) ports of rat-race coupler is crossed by the input and isolated ports. However, the performances of designed comparators are dropped due to non-equivalent group delay of modified couplers. Therefore, it is essential to design a comparator network achieving symmetrical input and output port allocation and reasonable transmission performance.

Additionally, to further mitigate the angular estimation error caused by amplitude & phase distortion that generated in the receiver, and noise, multi-path propagation interference from the measurement environment, the corresponding calibration and post correction processing are usually required in a monopulse receiving system, which needs extra hardware and more computation efforts. Previously, holographic back-

Manuscript received xxxxx. revised xxxxx; accepted xxxxx. This work is partially supported by National Science Foundation (NSF) under awards ECCS-2124531 and CCF-2124525. This paper is an expanded version from the IEEE Radio & Wireless Week (RWW), San Antonio, Texas, USA, 21-24 January 2024. (*Corresponding author*: Bayaner Arigong)

H. Zhang, S. Z. Pour, H. Yan, P. Liu, and B. Arigong is with Department of Electrical & Computer Engineering, College of Engineering, Florida A&M University - Florida State University, 2525 Pottsdamer St., Tallahassee, FL 32310 USA (e-mail: barigong@eng.famu.fsu.edu).

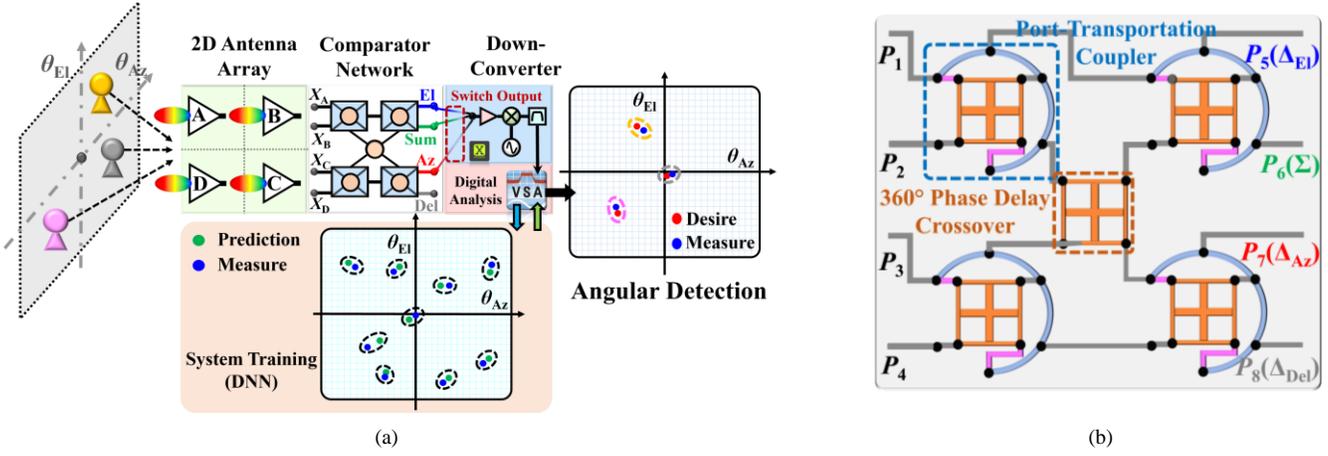

Fig. 1 The proposed low-cost monopulse receiver with enhanced DoA estimation accuracy via DNN real-time processing (a) descriptive diagram (b) proposed planar symmetrical monopulse comparator network.

projection method is proposed in [18] where it adds massive hardware to back of the radar for processing the signals. In [19], a multiple baseline interferometric radar is proposed to add an extra antenna in each dimension with a considerable distance from the primary array of antenna, which increases the estimation accuracy by sacrificing portability and simplicity. Recently, the deep neural network (DNN) learning, as one of the most outstanding techniques in artificial intelligence (AI) algorithms, attracts great interest in various applications. It enables automatic learning on complex patterns from a large amount of data, such that it extracts features from covariance matrix for constructing a nonlinear mapping [20]. Also, compared to a simple look-up table (LUT), DNN advances in its ability of reasonable prediction, handling noise and uncertainty appearing on the input data, and achieving a high memory efficiency. The deep learning-based classification of radar targets has been studied in [21], and a DNN for direction of arrival (DoA) estimation is proposed in [22], where the input is the covariance matrix to preserve part of the useful information. In [23], DNN is applied to fall detection with real-time feature extraction. [24] introduces DNN for DoA estimation enhancement with finite demonstration. So far, there is only limited study on merging DNN into a monopulse array to enhance the accuracy of angular estimation, especially for a low-cost radar system with simplified hardware integration and limited computation resources.

Our recent work, designing a planar symmetrical monopulse comparator network [25], is presented at 2024 IEEE Radio & Wireless Week, where it is composed of four port-transformation rat-race couplers and a 360° phase delay microwave crossover to overcome design challenges as unsymmetrical layout, amplitude/phase imbalance, and limited bandwidth. Expanding from this comparator network, in this extended paper, a low-cost high accuracy planar monopulse tracking array is developed by integrating antenna array, down conversion hardware, and DNN learning estimation. Here, our DNN learning algorithm requires small size of dataset to achieve fast training and high accuracy real-time tracking. To be specific, the followings are included in this extended paper:

1) more detailed design theory is derived for the fully planar symmetrical comparator network; 2) a monopulse receiver is designed by integrating a 2 × 2 patch antenna array, Keysight U3851A microwave down-converter as well as Keysight vector signal analysis (VSA) software; 3) intensive experiments are carried out to validate the $S$-parameter of comparator network and radiation characteristics of the monopulse array; 4) DNN learning algorithm is designed to accurately estimate DoA of moving target for the proposed monopulse tracking radar in different scenarios. In section II, the design theory is introduced for the proposed planar symmetrical comparator network. In section III, a prototype of the proposed monopulse receiver operating at 2 GHz is designed and tested and compared with the state-of-the-art other monopulse receivers. Also, the design of DNN learning algorithm is given and implemented for real-time physical location detection of moving object with high tracking accuracy. Finally, the future work is discussed in section IV.

II. DESIGN THEORY OF A FULLY SYMMETRICAL MONOPULSE COMPARATOR NETWORK

The schematic of proposed monopulse receiver system is depicted in Fig. 1 (a). It is composed of a four-element 2D antenna array, a fully symmetrical planar monopulse comparator network, a down-converter link including vector signal analysis software, and a deep neural network (DNN) learning algorithm with the training data and track the moving object location. In Fig. 1 (b), it shows the proposed comparator network which is composed of four identical new 180° couplers and a crossover with 360° phase delay to enable a fully planar topology with symmetrical input and output ports.

A. Novel Port Transformation 180° Rat-Race Coupler

The design flow of the proposed port transformation coupler is described in Fig. 2. The conventional 180° rat-race coupler features an unsymmetrical topology, where the sum (Σ) and delta (Δ) input ports ($P_a$ & $P_c$) are crossed by its output ports, as shown in Fig. 2 (a). To resolve this crossing issue, one of input ports $P_c$ is reallocated to a new position where it is at the same side of another input port $P_a$, as shown in Fig. 2 (b). Then, a microwave crossover is inserted into the conventional rat-race coupler to help transfer the port position, where 360° phase delay is added between the input and output of the crossover. To be more specific, a phase shifter $TL_\eta$ with impedance $Z_\eta$ and phase delay $\theta_\eta = 2\theta_1 = 180°$ (two pink lines) connects the crossover ports with the modified coupler from point ① to ② so that port transformation is realized as shown in Fig. 2 (b).

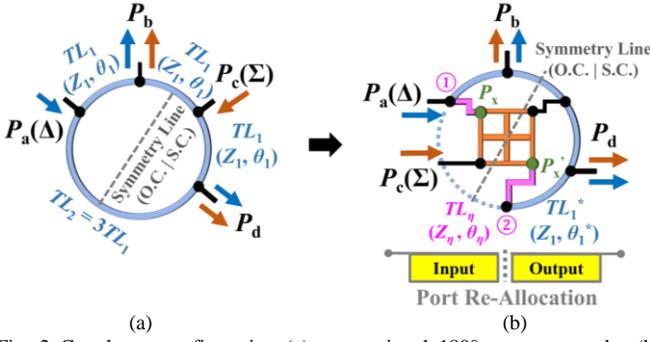

(a) (b)

Fig. 2 Coupler re-configuration (a) conventional 180° rat-race coupler (b) proposed symmetrical port transformation rat-race coupler.

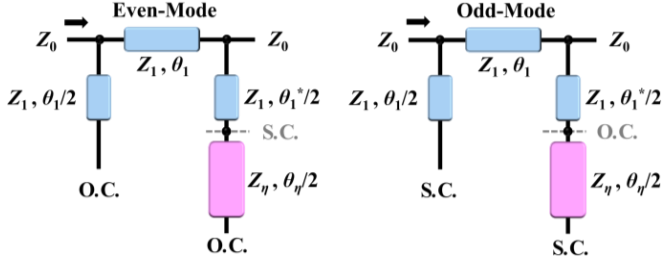

Fig. 3 The even-mode and odd-mode equivalent circuits model for proposed port-transformation rat-race coupler.

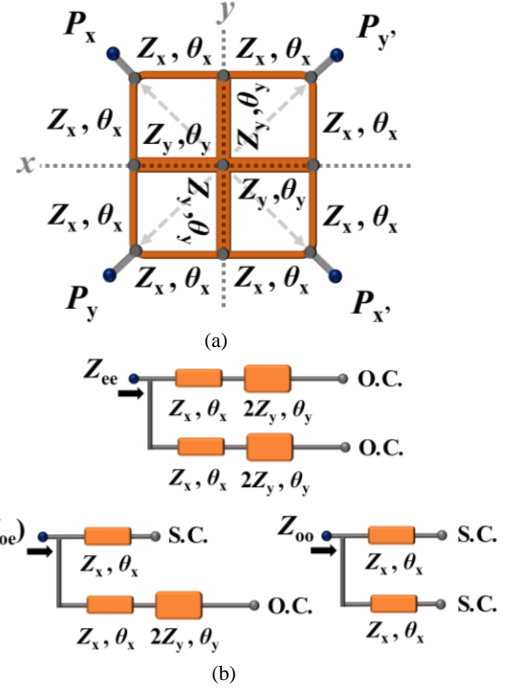

Fig. 4 360° phase delay microwave crossover (a) schematic diagram ($Z_x = 57\Omega$, $Z_y = 50\Omega$, $\theta_x = \theta_y = 90°$) (b) reduced networks with equivalent input under different excitations (i.e., even–even, even–odd, odd–even, and odd–odd).

The modified rat-race coupler is evenly split according to symmetry line as in Fig. 2 (b), where each section of the outer ring is composed of four identical quarter-wavelength transmission line sections $TL_1$ ($Z_1 = \sqrt{2} Z_0$, $\theta_1 = 90°$) and half of the phase shifter $TL_\eta$ ($Z_\eta$, $\theta_\eta$). The even-odd mode of equivalent circuits is given in Fig. 3, and the ABCD - matrix for even and odd mode can be derived as [26]:

$$\begin{bmatrix} A & B \\ C & D \end{bmatrix}_{even} = \begin{bmatrix} 1 & 0 \\ \dfrac{1}{-jZ_1 \cot \dfrac{\theta_1}{2}} & 1 \end{bmatrix} \begin{bmatrix} \cos\theta_1 & jZ_1 \sin\theta_1 \\ j\dfrac{1}{Z_1}\sin\theta_1 & \cos\theta_1 \end{bmatrix} \begin{bmatrix} 1 & 0 \\ \dfrac{1}{jZ_1 \tan \dfrac{\theta_1^*}{2}} & 1 \end{bmatrix} \quad (1)$$

$$= \begin{bmatrix} 1 & 0 \\ j/\sqrt{2}Z_0 & 1 \end{bmatrix} \begin{bmatrix} 0 & j\sqrt{2}Z_0 \\ j/\sqrt{2}Z_0 & 0 \end{bmatrix} \begin{bmatrix} 1 & 0 \\ -j/\sqrt{2}Z_0 & 1 \end{bmatrix}$$

$$\begin{bmatrix} A & B \\ C & D \end{bmatrix}_{odd} = \begin{bmatrix} 1 & 0 \\ \dfrac{1}{jZ_1 \tan \dfrac{\theta_1}{2}} & 1 \end{bmatrix} \begin{bmatrix} \cos\theta_1 & jZ_1 \sin\theta_1 \\ j\dfrac{1}{Z_1}\sin\theta_1 & \cos\theta_1 \end{bmatrix} \begin{bmatrix} 1 & 0 \\ \dfrac{1}{-jZ_1 \cot \dfrac{\theta_1^*}{2}} & 1 \end{bmatrix} \quad (2)$$

$$= \begin{bmatrix} 1 & 0 \\ -j/\sqrt{2}Z_0 & 1 \end{bmatrix} \begin{bmatrix} 0 & j\sqrt{2}Z_0 \\ j/\sqrt{2}Z_0 & 0 \end{bmatrix} \begin{bmatrix} 1 & 0 \\ j/\sqrt{2}Z_0 & 1 \end{bmatrix}$$

Specifically, in the even-odd mode equivalent circuits, it is found that the phase shifter with impedance $Z_\eta$ and electrical length $\theta_\eta/2 = 90°$ is either open or short. Therefore, its characteristic impedance $Z_\eta$ is not contributed to the ABCD - matrix in equation (1) and (2). In the other word, the impedance of this 180° phase shifter can be arbitrary chosen, which gives the design freedom to integrate the crossover into the coupler. With this derived feature, the 360° phase delay crossover with port impedance $Z_0 = 50 \ \Omega$ can physically move the original $P_c$ to the new $P_c$, as shown in Fig. 2 (b). Additionally, the electrical length of transmission lines (from $P_a$ to $P_x$, and from $P_x'$ to $P_d$) that connecting the 360° phase delay crossover and reconstructed coupler can be adjusted to satisfy $\theta_\eta = 180°$ and $\theta_1^* = \theta_1 = 90°$. Such that the group delay difference between components can be minimized and realize a decent bandwidth of the proposed port transformation 180° coupler.

B. *360° Phase Delay Microwave Crossover*

The microwave crossover is applied in proposed coupler to reallocate the input and output ports, and it has requirement to cause full wavelength delay from input ports to its output ports. Therefore, a crossover with 360° phase delay is developed and shown in Fig. 4 (a), where it is composed of the outer branch lines and inner crossed lines. The characteristic impedance and electrical length of transmission lines are denoted as ($Z_x$, $\theta_x$) and ($Z_y$, $\theta_y$). Since crossover is symmetric both in *x*- and *y*- planes, the even-odd mode analysis is applied to determine the transmission lines. The input impedances of a simplified network are denoted as $Z_{ee}$, $Z_{eo}$, $Z_{oe}$, and $Z_{oo}$ are shown in Fig. 4 (b). The corresponding reflection coefficients based on different excitations can be expressed as

$$\Gamma_{ee,eo,oe,oo} = \dfrac{Z_{ee,eo,oe,oo} - Z_0}{Z_{ee,eo,oe,oo} + Z_0} \quad (3)$$

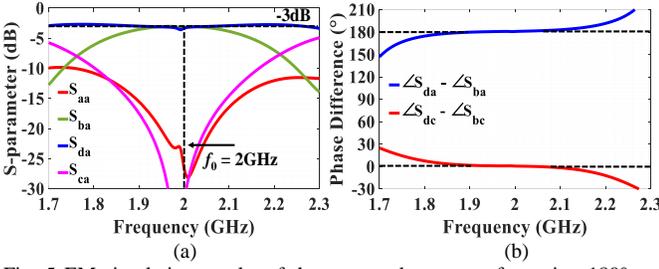

Fig. 5 EM simulation results of the proposed port-transformation 180° rat-race coupler: (a) magnitude response (b) phase difference for input delta (Δ) and sum (Σ) ports ($P_a$ & $P_c$) excitation.

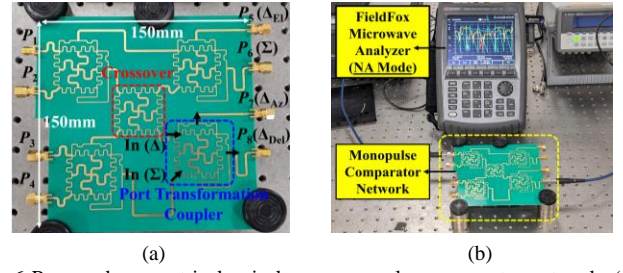

Fig. 6 Proposed symmetrical uni-planar monopulse comparator network: (a) PCB prototype circuit (b) measurement setup.

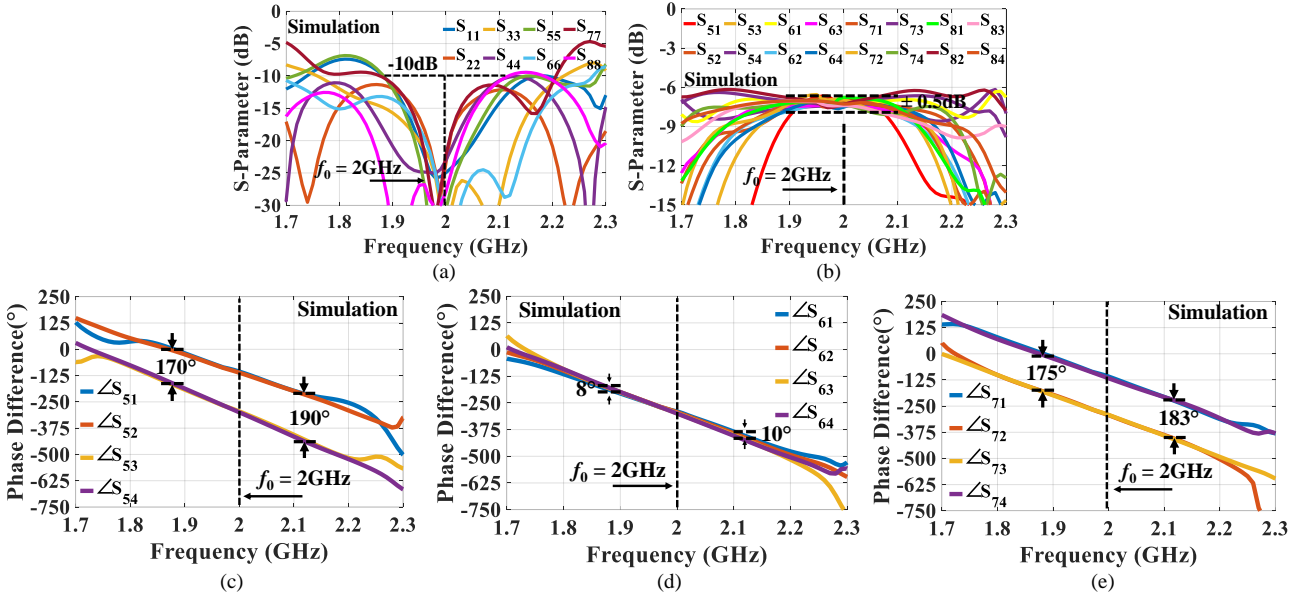

Fig. 7 EM simulations of proposed planar symmetrical comparator network: (a) reflections and isolations (b) transmissions (c) phase difference at azimuth port (d) phase differencet at sum port (e) phase difference at elevation port.

Where $Z_{ee}$, $Z_{eo}$, $Z_{oe}$, and $Z_{oo}$ can be derived from the theory in the previous work [27]. To achieve the function of the crossover, it should satisfy:

$$|S_{xx}| = |S_{xy'}| = |S_{xy}| = 0 \quad (4)$$

$$|S_{xx'}| = |S_{yy'}| = 1 \quad (5)$$

Thus, to meet the requirements in equation (4) and (5), the following equations are obtained:

$$\theta_y = 90° \quad (6)$$

$$(\tan\theta_x)^2 + 2\cdot\left[1-(\tan\theta_x)^2\right]\cdot\left(\frac{Z_0}{Z_x}\right) = 0 \quad (7)$$

$$\frac{2\cdot\sqrt{2\cdot(\tan\theta_x)^4 - 2\cdot(\tan\theta_x)^2}}{(\tan\theta_x)^3 - 3\cdot\tan\theta_x} = 0 \quad (8)$$

From equation (7) and (8), it is observed that there are multiple solutions existed to satisfy the condition of the crossover. To realize the 360° phase delay from input to output ports, the design parameters for inner and outer lines are selected as $Z_x = 57\ \Omega$, $Z_y = 50\ \Omega$, $\theta_x = \theta_y = 90°$, which can release the frequency sensitive phase compensation transmission lines. This crossover is also applied to interconnect two stage proposed coupler in comparator network.

## III. EXPERIMENTS RESULTS AND ANALYSIS

### A. Fully Symmetrical Monopulse Comparator Network

To verify the design concept, the EM simulation is conducted using Keysight Advanced Design System (ADS) software. Fig. 5 shows the simulated frequency responses of the proposed port-transformation rat-race coupler at central operating frequency 2 GHz. Correlating with the coupler in Fig. 2 (b), it is observed that the insertion loss ($|S_{ba}|$ & $|S_{da}|$) is 3.2 dB at operating frequency, and the return loss and isolation ($|S_{aa}|$ & $|S_{ca}|$) are more than 24 dB. The phase differences between output ports are 180° and 0° for sum (Σ) and delta (Δ) ports excitation, respectively. The simulated operating bandwidth is about 14% based on amplitude imbalance of ± 0.5 dB and phase imbalance of ± 10°. With this port-transformation coupler, a planar monopulse comparator network is designed by integrating four of the couplers and another crossover achieving symmetrical input / output ports and low amplitude and phase imbalance. The prototype of proposed comparator network is designed and fabricated using Rogers 4350B PCB (thickness = 0.76mm, $\varepsilon_r$ = 3.48, tan$\delta$ = 0.004) as shown in Fig. 6 (a). The total dimension of the circuit is 150mm × 150mm (1λ × 1λ), and the symmetrical input ($P_1$ - $P_4$) & output ($P_5$ - $P_8$) ports are naturally located on two sides, respectively. The experimental set up is shown in Fig. 6 (b), where Keysight Fieldfox microwave analyzer (Network Analyzer mode) is applied to measure the scattering parameters.

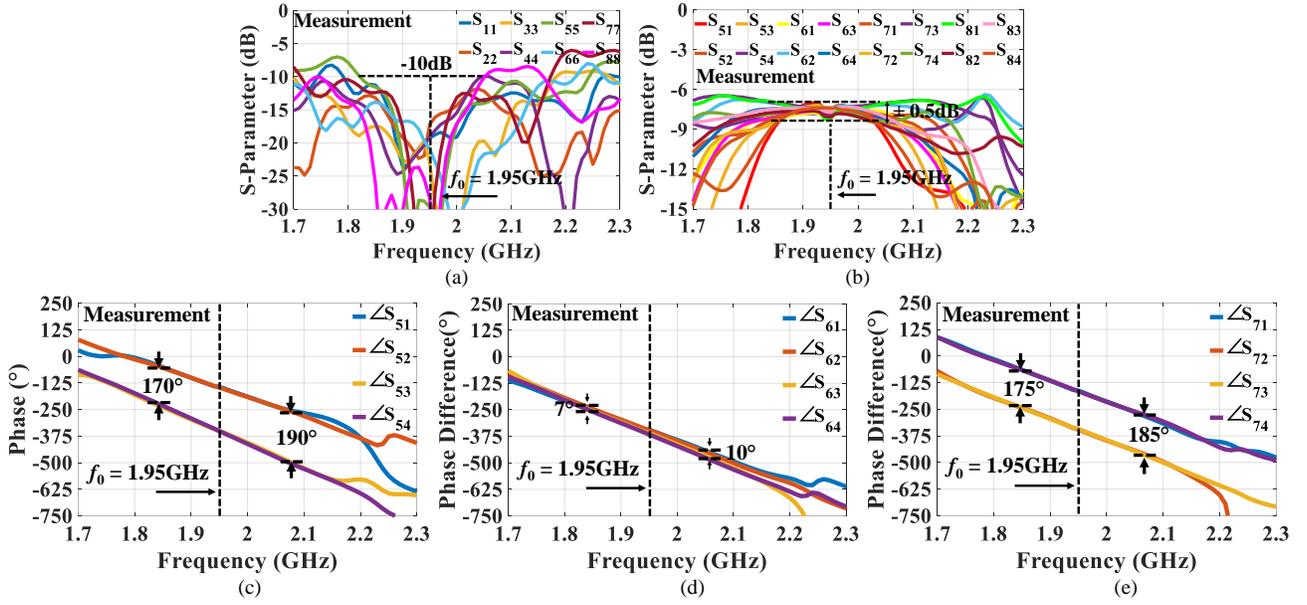

Fig. 8 Measurements of proposed planar symmetrical comparator network: (a) refelections and isolations (b) transmissions (c) phase difference at azimuth port (d) phase differencet at sum port (e) phase difference at elevation port.

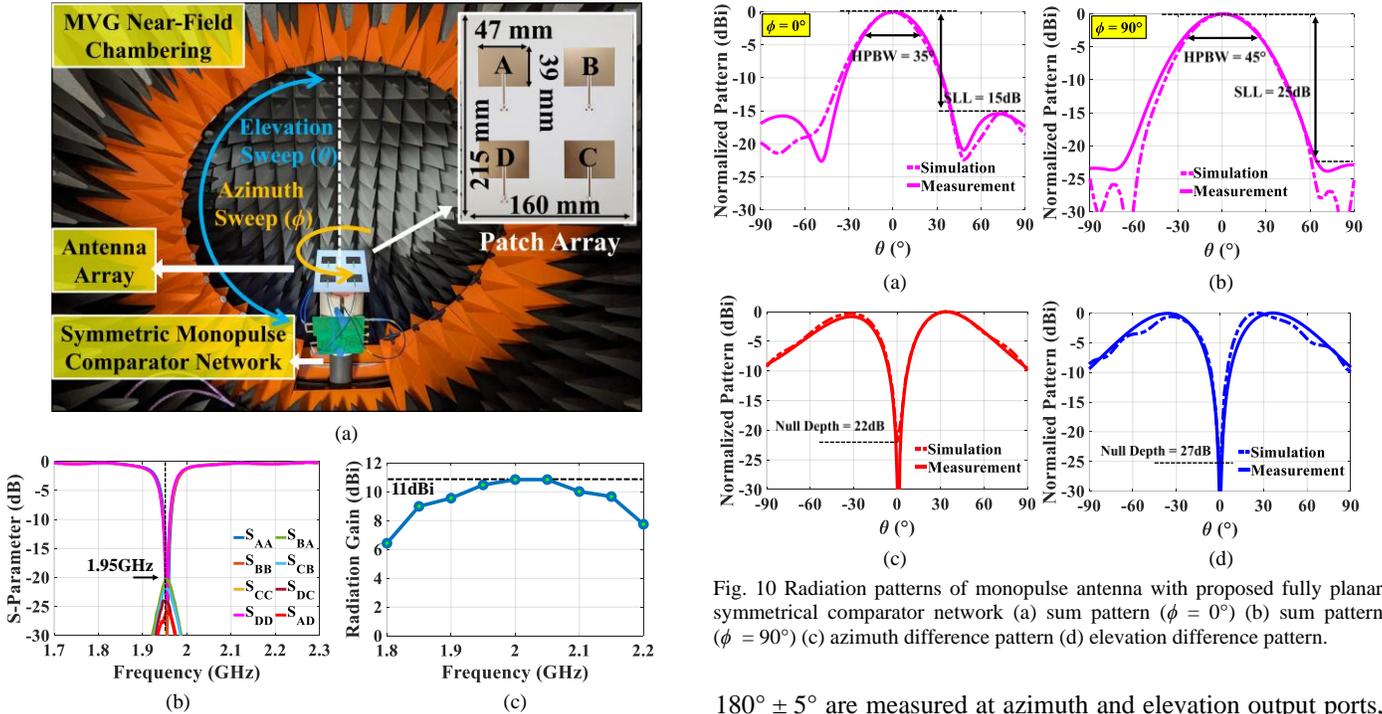

Fig. 9 Evaluation of the proposed monopulse array: (a) setup in MVG Starlab near-field chambering system (b) measured reflections and mutual couplings of the four-element antenna array (c) measured radiation gain of the patch antenna array.

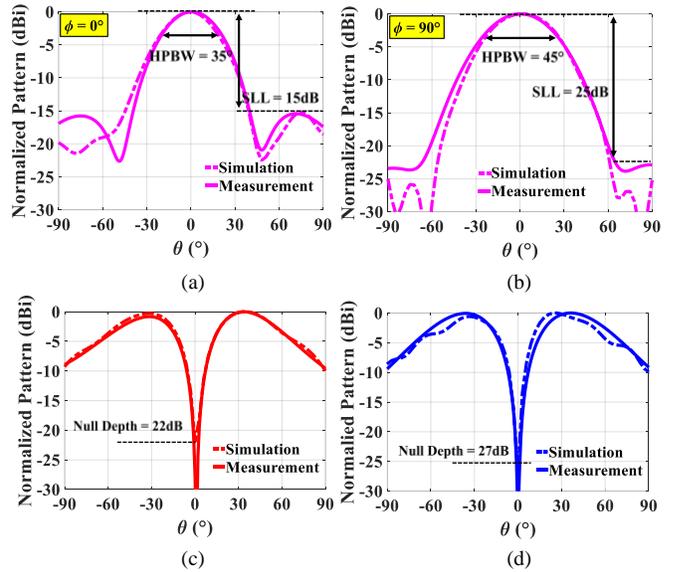

Fig. 10 Radiation patterns of monopulse antenna with proposed fully planar symmetrical comparator network (a) sum pattern ($\phi = 0°$) (b) sum pattern ($\phi = 90°$) (c) azimuth difference pattern (d) elevation difference pattern.

The simulated and measured results of the comparator network are shown in Fig. 7 and Fig. 8, respectively. In simulation, the transmission achieves 6.8 ± 0.5 dB at 2 GHz, and the return losses are better than 23 dB as plotted in Fig. 7 (a) - (b). In Fig. 7 (c) - (e), it shows the relative phase difference at different output ports where 0° ± 5° and 180° ± 5° are measured at azimuth and elevation output ports, respectively. The simulated operating bandwidth of proposed comparator network is about 12% (240 MHz) based on reflection & isolations of 10 dB, amplitude imbalance of ± 0.5 dB and phase imbalance of ± 10°. In the measurement, the transmission achieves 7.2 ± 0.5 dB at 1.95 GHz, and return losses are better than 17dB as plotted in Fig. 8 (a) - (b). The phase responses are shown in Fig. 8 (c) - (e), where 0° ± 5° and 180° ± 5° are achieved at azimuth and elevation output ports, respectively. The measured bandwidth of the comparator network is about 10% (195 MHz) based on reflection & isolations of 10 dB, amplitude imbalance of ± 0.5 dB and phase imbalance of ± 10°. The minor frequency shift between the simulation and measurement results mainly attribute to the fabrication tolerance.

TABLE I COMPARISON OF STATE-OF-THE-ART ON MONOPULSE ARRAY

| Reference | [4] | [6] | [11] | [14] | [16] | [17] | **This work** |
|---|---|---|---|---|---|---|---|
| Process Type | Multi-layer | Planar | SIW | SIW | 2D Conical | Planar | **Planar** |
| Symmetry | No | No | No | No | No | Yes | **Yes** |
| Element Type | Rat-race coupler | Branch-line coupler | Branch-line coupler | TEM line | Rat-race coupler | Symmetric coupler | **Symmetric coupler** |
| Network Size ($\lambda^2$) | 1.6 × 1.8 | 6.5 × 6 | 3.5 × 4.1 | 4 × 2 | 3 × 1.7 | 1.95 × 1.5 | **1 × 1** |
| Magnitude Imbalance | N.A. | 0.84 dB | 1 dB | N.A. | N.A. | 0.5 dB | **0.5 dB** |
| Phase Imbalance | N.A. | ± 2.5° | ± 10° | N.A. | N.A. | ± 10° | **± 10°** |
| Antenna Array Type | Patch (4) | Patch (192) | Patch (320) | Slot (12) | Conformal (4) | Patch (4) | **Patch (4)** |
| Aperture ($\lambda^2$) | 1.4 × 1.56 | 13 × 12 | 16.8 × 15.7 | 4 × 2 | N.A. | 1.95 × 1.5 | **1.1 × 1.4** |
| Gain (dBi) | 12.5 | 18.9 | 25.3 | 16.4 | 11 | N.A. | **11** |
| Mutual Coupling (dB) | -13 | -30 | -23 | -33 | N.A. | N.A. | **-20** |
| Side-Lobe Level (dB) | 19 | 17 | 24 | 9 | 15 | 15 | **15** |
| Null Depth (dB) | 27 | 30 | 30 | 29 | 26 | 20 | **22** |
| BW (%) | 2.1 | 5.6 | 2.8 | 100 | 22.2 | 8 | **10** |

*B. Monopulse Array Simulation and Measurement Results*

To further demonstrate the performance of the monopulse array designed from proposed fully symmetrical comparator network, a four-element 2 × 2 back-feed patch antenna array operating at 1.95 GHz is designed using the same PCB substrate as shown in Fig. 9. The size of the patch antenna array is 160 mm × 215 mm, where each antenna element is 39 mm × 47 mm. Compared to other types of directional antennas, such as Vivadi antenna, Yagi-Uda antenna, patch antenna achieves better uniformity on its radiation pattern in benefit of two-dimensional angular position detection, a good scalability in form of monopulse antenna array for better gain, null depth, half-power beamwidth, and it is easy to realize the optimization. The radiation characteristic of monopulse array is measured in the MVG StarLab near-field chamber system as shown in Fig. 9 (a). The individual reflections of four-element patch array are better than 13 dB at the operating frequency, and their mutual couplings are better than 20 dB, as shown in Fig. 9 (b). In Fig. 9 (c), the radiation gain of entire antenna array is up to 11 dBi.

The experimental results of sum and difference patterns ($\Delta_{Az}$ and $\Delta_{El}$) are plotted in Fig. 10, where a good agreement has been achieved between simulations and measurements. Specifically, Fig. 10 (a) & (b) show the sum patterns. The half-power beam width (HPBW) is 35° and 45° for $\phi = 0°$ and 90°, and the side-lobe levels (SLL) are better than 15 dB and 25 dB, respectively. The difference pattern in azimuth and elevation planes are shown in Fig. 10 (c) & (d), where null depths (ND) located at $\theta = 0°$ are better than 22 dB and 27 dB, respectively. The slight discrepancies between simulation and measurement in azimuth and elevations planes are attributed from minor variations from circuit fabrication and SMA connectors. Here, Table I compares the performance of the proposed monopulse array with state-of-the-art monopulse arrays, and it is found that our proposed comparator network features planar layout, symmetric, and compact size.

*C. DoA Estimation in Low-Cost Monopulse Receiver via DNN*

From the monopulse radar theory shown in Fig. 1, the coordinate information along azimuth (Az) and elevation (El)

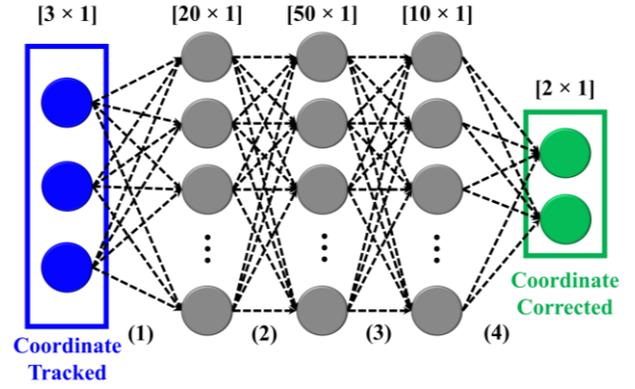

Fig. 11 The architecture of deep learning neural network implemented in the proposed low-cost monopulse receiver.

planes can be obtained after the received signals from antenna array ($X_A$, $X_B$, $X_C$, $X_D$) pass through the monopulse comparator network, and they are derived as:

$$\Delta_{Az} = (X_A + X_C) - (X_B + X_D) \quad (9)$$

$$\Delta_{El} = (X_A + X_B) - (X_C + X_D) \quad (10)$$

$$\Delta_{Del} = (X_A + X_D) - (X_B + X_C) \quad (11)$$

$$\Sigma = (X_A + X_B + X_C + X_D) \quad (12)$$

The DoA can be estimated from the monopulse ratio $\gamma$ that it is equivalent to the two-dimensional difference signals over the sum signal, which are derived as:

$$\gamma_{Az} = \frac{\Delta_{Az}}{\Sigma} = \tan\left[\frac{\pi d}{\lambda} \cdot \sin(\theta_{Az})\right] \quad (13)$$

$$\gamma_{El} = \frac{\Delta_{El}}{\Sigma} = \tan\left[\frac{\pi d}{\lambda} \cdot \sin(\theta_{El})\right] \quad (14)$$

Where $d$ is the distance between each antenna element of a monopulse array, and $\lambda$ denotes the wavelength of operating frequency. The angular information in azimuth ($\theta_{Az}$) and elevation ($\theta_{El}$) planes can be estimated from equations (13) and (14). However, there will be misalignment errors on the practical angular estimation due to amplitude and phase distortions that attribute from the array beamwidth, multi-path wireless transmission, and noise interference of the down-generates the corrected elevation and azimuth angles at the output. The implemented DNN consists of three consecutives fully connected hidden layers that they contain 20, 50, and 10 hidden neurons, respectively. For the training process, the weights, and biases in hidden layers of DNN are appropriately optimized to minimize the loss function with *N* data samples. Here, the loss function of the DNN is defined as:

$$L_{DNN} = \frac{1}{N}\sum_{i=1}^{N}\left[(\theta_{El\_DNN} - \theta_{El\_truth})^2 + (\theta_{Az\_DNN} - \theta_{Az\_truth})^2\right] \quad (15)$$

where $\theta_{El\_truth}$ and $\theta_{El\_DNN}$ are the elevation angles of the true coordinates and the angles predicted by the DNN, respectively. Similarly, $\theta_{Az\_truth}$ and $\theta_{Az\_DNN}$ are the azimuth angles of the true coordinates and the angles predicted by the DNN, respectively.

The angular estimation testbed is shown in Fig. 12 (a), where a standard patch antenna performing as the detected object is set in far-field range (*D*) from the monopulse receiver. The training dataset DNN contains 145 samples of continuous locations of the remote target (standard patch antenna), which is moving along 2D plane with unit of 30mm. More specifically, the true coordinate and predicted coordinates of the remote target in reference to monopulse receiver boresight, as denoted in equation (15), are recorded as two sets of parameters during the training process. 90% of the data set is allocated for training, and the remaining 10% serve as a validation. The azimuth and elevation estimations are carried out simultaneously via one DNN because the correlation of two planes, where the two-coordinate data are added to both the input and output of the DNN. Additionally, applying a single DNN simplifies the system, as it estimates the coordination information with less computational effort compared to using two separate DNNs to predict both elevation and azimuth misalignments. As shown in Fig. 12 (a), the monopulse receiver is composed by our proposed planar monopulse array and a Keysight U3851A microwave down-conversion. In specific, the monopulse array captures the RF signals ($\Delta_{Az}$, $\Delta_{El}$, and $\Sigma$), and the received RF signals are down converted into baseband signals at 100 MHz by switching the connection between each output port of comparator network and down-converter module. Then, Keysight vector signal analysis (VSA) software is applied to analyze the two-dimensional angular information.

In the experiments, the standard patch antenna target is moving along both azimuth and elevation axis, and the angular information is measured and estimated through the equations (13) and (14). The DNN training process takes about 6 minutes using a laptop with Intel Core™ i7-12700H Processor with 20,000 of iterations. Here, we use Adam optimizer for updating weights and biases of the DNN. After it is fully trained, the validation error is about $2 \times 10^{-6}$ that equivalent to a position error of 1.1 mm in a 2D plane with a 0.62 m separation distance between the transmitter target and the monopulse receiver. Fig. 12 (b) - (c) show the angular estimation results, where multiple samples are detected on four quarters of a spatial coordinate with two different detection distances (*D* = 0.66 m & 0.86 m). Compared to the angular information under initial tracking (blue), the coordinates processed by DNN (green) achieve a great converter link. To minimize the systematic degradation and mitigate the angular estimation error, a DNN learning processing is developed and implemented.

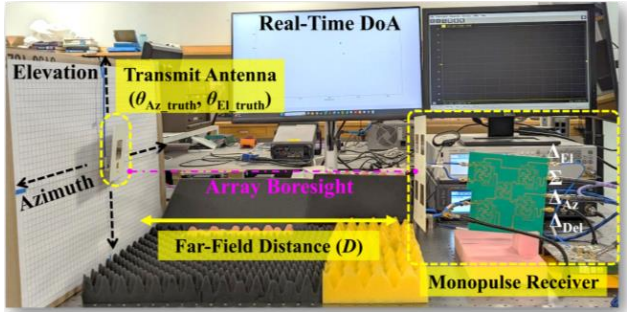
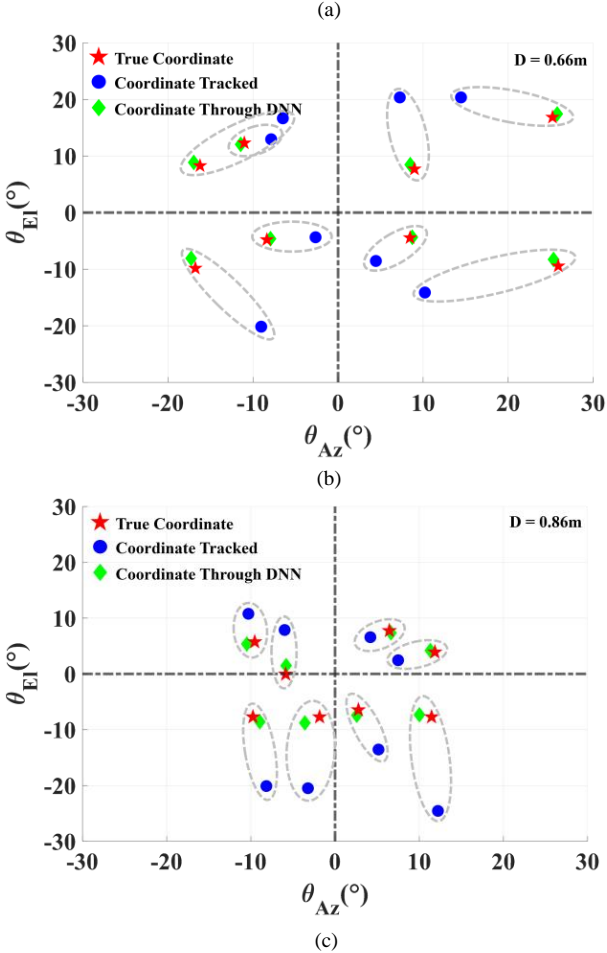

Fig. 12 DoA estimation using the proposed low-cost monopulse receiver system. (a) system testbed setup (b) angular estimation results with detection distance *D* = 0.66 m (c) angular estimation results with detection distance *D* = 0.86 m.

The proposed DNN aims to find the accurate coordinate of a target from the data processed by the proposed monopulse receiver and to achieve real-time monitoring of coordination data with minimum delay. In specific, the input data for the neural network is the elevation ($\theta_{El}$) and azimuth ($\theta_{Az}$) angles of the detected target that estimated by the proposed monopulse receiver. Moreover, the distance between the radar and the detected object is also included into the input data of the DNN. In Fig. 11, the neural network initiates with a 3-by-1 matrix as the input layer from the monopulse receiver, and it correlation with true coordinate of the target (red). Once the DNN is completely trained, it can effectively calibrate system degradation and some potential interferences, thereby providing accurate locations of detected targets in milliseconds enabling a real-time monopulse tracking radar system. In a summary, several investigations can be conducted to further improve the proposed monopulse receiver in the future work: 1). wideband coupler and crossover with minimum group delay can be designed to further improve the operating bandwidth of entire comparator network; 2). more antenna elements can be included in each array quarter to enhance the analog detection accuracy; 3). the wideband antenna array can be designed integrated on the backside to make the receiver more compact; 4). To achieve less ambiguity of DoA results, DNN can be improved by introducing additional input features and modifying network architecture.

## IV. Conclusion

In this paper, a low-cost monopulse receiver with an enhanced estimation on DoA has been presented, in which a fully symmetrical monopulse comparator network and DNN post-processing algorithm have been introduced to untether the complicated hardware design and to minimize the systematic degradation, respectively. Specifically, the proposed monopulse comparator network is composed of a port-transformation 180° rat-race coupler and a 360° phase delay crossover. It reallocates one of the input ports to resolve the output ports crossing issue in conventional rat-race coupler and achieve planar symmetrical topology of the monopulse comparator feeding network. To validate the design concept, a comparator network and monopulse array operating at 1.95 GHz are designed, fabricated, and verified in experiment. The measurement results align well with the simulation results for both *S*-parameters and radiation patterns. Additionally, a prototype testbed merged with DNN is designed to demonstrate the effectiveness of proposed system. Different from any other tracking radars, the proposed design can enable a real-time monitoring of angular information of the remote target using small amount of data samples.

## References

[1] P. Molchanov, S. Gupta, K. Kim, K. Pulli, "Short-Range FMCW Monopulse Radar for Hand-Gesture Sensing," *IEEE Radar Conference*, Arlinton, VA, USA, 2015, pp. 1491 – 1496.
[2] Z. Wang, Y. Hu, L. Xiang, J, Xu, W. Hong, "A Wideband High-Gain Planar Monopulse Array Antenna for Ka-Band Radar Applications," *IEEE Trans. Antennas Propag.*, vol. 71, no. 11, pp. 8739-8752, Nov. 2023.
[3] R. Feghhi, R. S. C. Winter, F. M. Sabzevari and K. Rambabu, "Design of a Low-Cost UWB Time-Domain Radar System for Subcentimeter Image Resolution," *IEEE Trans. Microw. Theory Tech*, vol. 70, no. 7, pp. 3617-3628, July. 2022.
[4] S. A. Khatami, J. Meiguni, A. Amn-e-Elahi, and P. Rezaei, "Compact Via-Coupling Fed Monopulse Antenna with Orthogonal Tracking Capability in Radiation Pattern," *IEEE Antennas Wireless Propag. Lett.*, vol. 19, no. 8, pp. 1443–1446, Aug. 2020.
[5] H. Kumar, G. Kumar, "Broadband Monopulse Microstrip Antenna Array for X-Band Monopulse Tracking," *IET Microw. Antennas & Propag.*, vol. 12, no. 13, pp. 2109-2114, Aug. 2018.
[6] H. Wang, D.-G. Fang, and X. G. Chen, "A Compact Single Layer Monopulse Microstrip Antenna Array," *IEEE Trans. Antennas Propag.*, vol. 54, no. 2, pp. 503–509, Feb. 2006.
[7] A. E. Tan, M. Y. Chia, K. Rambabu, "Design of Ultra-Wideband Monopulse Receiver," *IEEE Trans. Microw. Theory Tech*, vol. 54, no. 11, pp. 3821-3827, Nov. 2006.
[8] S. Alamdar, K. M. Aghdam, H. Khalili, "A Compact Monopulse Antenna Array with Suppressed Mutual Coupling Using Broadband Schiffman Phase Shifters," *IEEE Access* vol. 12, pp. 11936-11944, Jan. 2024.
[9] K. S. Ang, Y. C. Leong, and C. H. Lee, "A Wide-Band Monopulse Comparator with Complete Nulling in All Delta Channels Throughout Sum Channel Bandwidth," *IEEE Trans. Microw. Theory Techn.*, vol. 51, no. 2, pp. 371–373, Feb. 2003.
[10] H. Chung, Q. Ma, G. M. Rebeiz, "28GHz Active Monopulse Networks with Amplitude and Phase Control and -30dB Null-Bandwidth of 5GHz," *IEEE MTT-S International Microwave Symposium* (*IMS*), Los Angeles, CA, USA, 2020, pp. 1-4.
[11] L. Zou, X. Wang, and J. Zang, "Series-Fed Monopulse Microstrip Array Antenna with Stripline Quadrature Hybrid Comparator Network," *IEEE Access*, vol. 9, pp. 169177–169192, Dec. 2021.
[12] H. Chen, W. Che, Q. He, W. Feng, X. Wei and K. Wu, "Compact Substrate Integrated Waveguide (SIW) Monopulse Network for Ku - Band Tracking System Applications," *IEEE Trans. Microw. Theory Techn*, vol. 62, no. 3, pp. 472-480, Mar. 2014.
[13] Y. Cao, S. Yan, "A Dual-Mode SIW Compact Monopulse Comparator for Sum and Difference Multibeam Radar Applications," *IEEE Microw. Wireless Compon. Lett.*, vol. 32, no. 1, pp. 41-44, Jan. 2022.
[14] W. Li, S. Liu, J. Deng, Z. Hu and Z. Zhou, "A Compact SIW Monopulse Antenna Array Based on Microstrip Feed," *IEEE Antennas Wireless Propag. Lett.*, vol. 20, no. 1, pp. 93-97, Jan. 2021.
[15] D. Nagaraju and Y. K. Verma, "A compact wideband planar magic tee for monopulse antenna array applications," *IEEE Microw. Wireless Compon. Lett.*, vol. 31, no. 5, pp. 429–432, May 2021.
[16] Y. Gao, W, Jiang, W, Hu, Q. Wang, W, Zhang, and S. Gong, "A Dual-Polarized 2-D Monpulse Antenna Array for Conical Conformal Applications," *IEEE Trans on Antennas and Propag.*, vol. 69, no. 9, pp. 5479- 5499, Sept. 2022.
[17] H. Zhang, H. Ren, Y. Gu and B. Arigong, "A Fully Symmetrical Uni-Planar Microstrip Line Comparator Network for Monopulse Antenna," *IEEE Microw. Wireless Technol. Lett*, vol. 33, no. 5, pp. 611-614, May 2023.
[18] D. Schvartzman, J. D. D. Díaz, D. Zrnić, M. Herndon, M. B. Yeary and R. D. Palmer, "Holographic Back-Projection Method for Calibration of Fully Digital Polarimetric Phased Array Radar," *IEEE Trans. Radar Syst.*, vol. 1, pp. 295-307, Jun. 2023.
[19] J. Merlo, E. Klinefelter, S. Vakalis and J. A. Nanzer, "A Multiple Baseline Interferometric Radar for Multiple Target Angular Velocity Measurement," *IEEE Microw. Wireless Compon. Lett.*, vol. 31, no. 8, pp. 937-940, Aug. 2021.
[20] S. Xu, B. Chen, H. Xiang. "A Low Computational Complexity DOA Estimation Using Sum/Difference Pattern Based on DNN." *Multidimensional Systems and Signal Processing*, vol. 34, no. 1, July 2021.
[21] X. Cai, M. Giallorenzo and K. Sarabandi, "Machine Learning-Based Target Classification for MMW Radar in Autonomous Driving," *IEEE Trans. Intelligent Vehicles,* vol. 6, no. 4, pp. 678-689, Dec. 2021.
[22] L. Wu, Z. -M. Liu and Z. -T. Huang, "Deep Convolution Network for Direction of Arrival Estimation with Sparse Prior," *IEEE Signal Processing Lett.*, vol. 26, no. 11, pp. 1688-1692, Nov. 2019.
[23] B. Jokanović and M. Amin, "Fall Detection Using Deep Learning in Range-Doppler Radars," *IEEE Trans. Aerospace and Electronic Syst.*, vol. 54, no. 1, pp. 180-189, Feb. 2018.
[24] H. Zhang, et al., "Boosting Estimation Accuracy of Low-Cost Monopulse Receiver Via Deep Neural Network," *2024 IEEE Wireless and Microwave Technology Conference (WAMICON)*, Clearwater, FL, USA, April 2024, pp. 1-4.
[25] H. Zhang, P. Liu, J. Casamayor, S. Z. Pour, M. Plaisir, B. Arigong, "A Planar Monopulse Comparator Network Design from Port-Transformation Rat-Race Coupler," *2024 IEEE Radio & Wireless Week* (*RWW*), San Antonio, TX, USA, Jan. 2024, pp. 1-4.
[26] D. M. Pozar, Microwave Engineering, 3rd ed. New York, NY, *USA: Wiley*, 2005.
[27] H. Ren, M. Zhou, H. Zhang and B. Arigong, " A Novel Dual-Band Zero-Phase True Crossover with Arbitrary Port Impedance," *IEEE Microw. Wireless Comp. Lett*, vol. 29, no. 1, pp. 29-31, Jan. 2019.